\documentclass[twocolumn,showpacs]{revtex4}

\usepackage{graphicx}
\usepackage{dcolumn}
\usepackage{amsmath}
\makeatletter
\def\btt#1{\texttt{\@backslashchar#1}}
\DeclareRobustCommand\bblash{\btt{\@backslashchar}}
\makeatother

\begin{document}

\title{Horizons of radiating black holes in Einstein gauss-bonnet gravity}

\author{S. G.~Ghosh}
\email{sgghosh@iucaa.ernet.in} \affiliation{BITS, Pilani DUBAI
Campus, P.B. 500022, Knowledge Village, DUBAI, UAE  and
Inter-University Center for Astronomy and Astrophysics,
 Post Bag 4 Ganeshkhind,  Pune - 411 007, India }
\author{D. W.~Deshkar}
\affiliation{S.S.E.S. Amti's Science College, Congress Nagar,
Nagpur-440 012, India}

\date{\today}

\begin{abstract}
A Vaidya-based model of a radiating black-hole is studied in a 5-dimensional
Einstein gravity with Gauss-Bonnet contribution of quadratic curvature terms.
 The structure and locations of the apparent and event horizons of the radiating black hole are determined.
\end{abstract}


\maketitle

\section{Introduction}
The Vaidya metric \cite{pcgb}, which has the form
\begin{equation}\label{vm}
ds^2 = - \left[1 - \frac{2 m(v)}{r}\right] d v^2 + 2 \epsilon d v
d r + r^2 (d \theta^2+ sin^2 \theta d \phi^2), \;
\epsilon \pm 1
\end{equation}
is a solution of Einstein's equations with spherical symmetry for
a null fluid (radiation) source described by energy momentum
tensor  $T_{ab} = \mu l_a l_b$, $ l_a$ being a null vector
field. For the case of an ingoing radial flow, $ \epsilon = 1$ and
$m(v)$ is a monotone increasing mass function in the advanced time
$v$, while $\epsilon = -1$ corresponds to an outgoing radial flow,
with $m(v)$ being in this case a monotone decreasing mass function
in the retarded time $v$. Also, several solutions in
which the source is a mixture of a perfect fluid and null radiation
have been obtained in later years \cite{kagb,bvgb,vhgb,gkgb,hcgb,wwgb,dggb}.
The Vaidya-based  metric is today commonly used for
 the study of Hawking radiation, the process of black-hole evaporation \cite{rp1gb}
 and study of the dynamical evolution of
 the horizon associated with radiating black holes \cite{jygb,rlm,km,bdk,mrm}.

In recent years, motivated by development in the string theory,
there has been renewed interest in the theories of gravity in higher
dimensions. As a possibility the Einstein-Gauss-Bonnet gravity is
low energy limit of the string theory is of particular interest
because of it's special features. In this paper, we consider the
5D  action with the Gauss-Bonnet terms for gravity and
give a model of the gravitational collapse of a null fluid
including the perturbative effects of quantum gravity. The
Gauss-Bonnet terms are the higher curvature corrections to general
relativity and naturally arise as the next leading order of the
$\alpha$-expansion of heterotic superstring theory, where $\alpha$
is the inverse string tension \cite{Gross}.

 The aim of this brief report is to study how the location, character and evolution of event horizon (EH)
 and apparent horizon (AH) get modified in the presence of Gauss-Bonnet term.
 The calculations are  based on the recently introduced exact solution for a radiating Vaidya solution in
 the Einstein-Gauss-Bonnet gravity \cite{hm,tk}.

\section{Vaidya Solution in Einstein Gauss-Bonnet Gravity}
We begin with a review of Einstein Gauss-Bonnet gravity and Vaidya radiating black hole solution in it.
The gravitational part of the  $5$-dimensional ($5D$) action that we
consider is:
\begin{equation}
\label{action} S=\int
d^5x\sqrt{-g}\biggl[\frac{1}{2\kappa_5^2}(R-2\Lambda+\alpha{L}_{GB})
\biggr]+S_{\rm matter},
\end{equation}
where $R$ and $\Lambda$ are the 5D Ricci scalar and the cosmological
constant, respectively. $\kappa_5\equiv\sqrt{8\pi G_5}$, where $G_5$
is the 5D gravitational constant. The Gauss-Bonnet Lagrangian is the
combination of the Ricci scalar, Ricci tensor $R_{ab}$, and Riemann
tensor $R^a_{~~b\rho\sigma}$ as
\begin{equation}
{L}_{GB}=R^2-4R_{ab}R^{ab}+R_{ab\rho\sigma}R^{ab\rho\sigma}.
\end{equation}
In the 4-dimensional space-time, the Gauss-Bonnet terms do not
contribute to the field equations. $\alpha$ is the coupling constant
of the Gauss-Bonnet terms. This type of action is derived in the
low-energy limit of heterotic superstring theory~\cite{Gross}. In
that case, $\alpha$ is regarded as the inverse string tension and
positive definite and we consider only the case with $\alpha \ge 0$
in this paper. We consider a null fluid as a matter field,
whose action is represented by $S_{\rm matter}$ in
Eq.~(\ref{action}).

From the action (\ref{action}) we derive the following field
equations:
\begin{equation}
{G}_{a b} - \alpha {H}_{a b} =  {T}_{ab}, \label{beqgb}
\end{equation}
where
\begin{eqnarray}
{G}_{ab}&=&R_{ab}-{1 \over 2}g_{ab}R,\\
{H}_{ab}&=&2\Bigl[RR_{ab}-2R_{a \alpha}R^{\alpha}_{b}-2R^{\alpha
\beta}R_{a \alpha b\beta}
 +R_{a}^{~\alpha\beta\gamma}R_{b \alpha\beta\gamma}\Bigr]\nonumber\\
&&-{1\over 2}g_{ab}{L}_{GB}.
\end{eqnarray}
The energy-momentum tensor of a null fluid is
\begin{eqnarray}
{T}_{ab}=\mu l_{a}l_{b}, \label{eq:emtgb}
\end{eqnarray}
where $\mu$ is the non-zero energy density and $l_a$ is a null
vector such that
\begin{eqnarray}
l_{a} = \delta_a^0,\;  l_{a}l^{a} = 0.
\end{eqnarray}
Expressed in terms of Eddington advanced time coordinate (ingoing
coordinate) $v$, the metric of general spherically symmetric
space-time
\begin{equation}
ds^2 = - A(v,r)^2 f(v,r)\;  dv^2
 +  2 A(v,r)\; dv\; dr + r^2 d \Omega^2_3, \label{eq:me1gb}
\end{equation}
$ d\Omega^2_3 = d \theta^2+ \sin^2 \theta d \phi^2 + \sin^2 \theta
 \sin^2\phi^2 d\psi^2$. Here $A$ is an arbitrary function.
  We wish to find the general
solution of the Einstein equation for the matter field given by
Eq.~(\ref{eq:emtgb}) for the metric (\ref{eq:me1gb}), which contains
two arbitrary functions. It is the field equation $G^0_1 = 0$ that
leads to $ A(v,r) = g(v)$. This could be absorbed by writing $d
\tilde{v} = g(v) dv$.  Hence, without loss of generality, the metric
(\ref{eq:me1gb}) takes the form,
\begin{equation}
ds^2 = - f(v,r) d v^2 + 2 d v d r + r^2 d  \Omega^2_3,
\label{eq:megb}
\end{equation}
The Einstein field equations take the form:
\begin{equation}
f' - \frac{2}{r}(1-f)+ \frac{4\alpha}{r^2}(1-f)f'=0, \label{feqgb}\\
\end{equation}
\begin{equation}
f{'}' + \frac{4}{r}f' + \frac{2}{r^2} (1-f) + \frac{4
\alpha}{r^2}\left[ f{'}'(1-f) + f{'}^2 \right]=0, \\
\end{equation}
\begin{equation}
\mu  = \frac{3}{2}\frac{\dot{f}}{r} + \frac{6
\alpha}{r^3}\dot{f}(1-f).
\end{equation}
Then, $f$ is obtained by solving only the  (\ref{feqgb}),  This
equation is integrated to give the general solution as
\begin{equation}
f = 1 + \frac{r^2}{4 \alpha} \left[ 1 \pm \sqrt{1 + \frac{8 \alpha
m(v)}{r^4}}\right], \label{s-feqgb}
\end{equation}
where $m(v)$ is an arbitrary function of $v$. The
special case in which $m$ is a non-zero constant we call the
GB-Schwarzschild solution, of which the global structure is
presented in~\cite{tmgb}.

There are two families of solutions which correspond to the sign in
front of the square root in Eq.~(\ref{s-feqgb}). We call the family
which has the minus (plus) sign the minus- (plus+) branch solution.
From the $(r,v)$ component of (\ref{beqgb}), we obtain the energy
density of the null fluid as
\begin{eqnarray}
\mu=\frac{3}{2r^{3}}{\dot m}, \label{density}
\end{eqnarray}
for both branches, where the dot denotes the derivative with respect
to $v$. In order for the energy density to be non-negative, ${\dot
m} \ge 0$ must be satisfied.  In the general relativistic limit ${
\alpha} \to 0$, the minus-branch solution reduces to
\begin{eqnarray}
f=1-\frac{m(v)}{r^{2}}, \label{vaidyaGR}
\end{eqnarray}
which is the $5D$ Vaidya solution in Einstein gravity \cite{gs}. It may noted
that, in $5D$ Einstein gravity, the density is still given by
Eq.~(\ref{density}). There is no such limit for the plus-branch
solution. In the static case ${\dot m}=0$, this solution reduces to
the solution which was independently discovered by Boulware and
Deser~\cite{GB_BH} and Wheeler~\cite{Wheeler_1}.

The Kretschmann scalar ($\mbox{KS} = R_{abcd} R^{abcd}$, $R_{abcd}$ is the
5D Riemann tensor) and Ricci scalar ($\mbox{R} = R_{ab} R^{ab}$, $R_{ab}$
is the 5D Ricci tensor) for the metric (\ref{eq:megb}) reduces to
\begin{equation}
\mbox{KS} = f{''}^2+\frac{6}{r^4}f{'}^2 + \frac{12}{r^4}
(1-f)^2,\label{ksgb}
\end{equation}
and
\begin{equation}
\mbox{R} = f{''}+ \frac{6}{r}f'- \frac{6}{r^2} (1-f),\label{rsgb}
\end{equation}
which diverges at $r=0$ and hence the singularity is a scalar
polynomial \cite{hegb}.
Radial ($ \theta$ and $
\phi \,=\,const$.) null geodesics of the metric (\ref{eq:megb}) must
satisfy the null condition
\begin{equation}
2 \frac{dr}{dv} =1 + \frac{r^2}{4 \alpha} \left[ 1 \pm \sqrt{1 + \frac{8 \alpha
m(v)}{r^4}}\right], \label{eq:de1gb}
\end{equation}
The nature (a naked singularity or a black hole) of the collapsing solutions
can be characterized by the existence of radial null geodesics coming out from
 the singularity.  It has been shown that a time-like naked singularity is formed, which does not appear in the general
 relativistic case \cite{hm}.
\section{RADIATING BLACK HOLE HORIZONS}
In this section, we study the structure and location of the EH's and
AH`s in the presence in Einstein-Gauss-Bonnet gravity and compare it
with that in general relativity case by use of solution obtained in
the previous section.  We consider the minus-branch solution in
order to compare with general relativistic case. The line element of
the radiating black hole in Einstein-Gauss-Bonnet gravity has the
form (\ref{eq:megb}) with $f$ given by Eq.~(\ref{s-feqgb}) and  the
energy momentum tensor (\ref{eq:emtgb}). The luminosity due to loss
of mass is given by $L_M = - dM/dv$, $L_M < 1$ measured in the
region where $d/dv$ is time-like. In order to further discuss the
physical nature of our solutions, we introduce their kinematical
parameters. Following York \cite{jygb} a null-vector decomposition
of the metric (\ref{eq:megb}) is made of the form
\begin{equation}\label{gabgb}
g_{ab} = - \beta_a l_b - l_a \beta_b + \gamma_{ab},
\end{equation}
where,
\begin{eqnarray}
\beta_{a} &=& - \delta_a^v, \: l_{a} = - \frac{1}{2} f(v,r)
\delta_{a}^v + \delta_a^r, \label{nvagb}
 \\
\gamma_{ab} &=& r^2 \delta_a^{\theta} \delta_b^{\theta} + r^2
\sin^2(\theta) \delta_a^{\varphi} \delta_b^{\varphi}
 \nonumber\\&&+ r^2 \sin^2(\theta)\sin^2(\phi) \delta_a^{\psi} \delta_b^{\psi},
\label{nvbgb}
\\
l_{a}l^{a} &=& \beta_{a} \beta^{a} = 0, \; ~l_a \beta^a = -1,\; ~l^a
\;\gamma_{ab} = 0, \nonumber\\&& ~\gamma_{ab}\; ~\beta^{b} = 0,
\label{nvdgb}
\end{eqnarray}
 with $f(v, r)$ given by Eq.~(\ref{s-feqgb}).  The optical behavior
of null geodesics congruences is governed by the Raychaudhuri
equation
\begin{equation}\label{regb}
   \frac{d \Theta}{d v} = K \Theta - R_{ab}l^al^b-\frac{1}{2}
   \Theta^2 - \sigma_{ab} \sigma^{ab} + \omega_{ab}\omega^{ab},
\end{equation}
with expansion $\Theta$, twist $\omega$, shear $\sigma$, and surface
gravity $K$. The expansion of the null rays parameterized by $v$ is
given by
\begin{equation}\label{theta}
\Theta = \nabla_a l^a - K,
\end{equation}
where the $\nabla$ is the covariant derivative and the surface
gravity is
\begin{equation}\label{sggb}
K = - \beta^a l^b \nabla_b l_a.
\end{equation}
The AH is the outermost marginally trapped
surface for the outgoing photons.  The AH can be either null or
space-like, that is, it can 'move' causally or acausally.  The
apparent horizons are defined as surface such that $\Theta \simeq 0$
which implies that $f=0$. Using Eqs.~(\ref{s-feqgb}), (\ref{nvagb}) and (\ref{sggb})
\begin{equation}
K = \frac{r}{4 \alpha} \left[1 - \sqrt{1 + 8 \alpha \frac{m(v)}{r^4}
}\right] + \frac{\frac{2m(v)}{r^3} }{\sqrt{1 + 8 \alpha
\frac{m(v)}{r^4} }}.\label{Kgb}
\end{equation}
Then Eqs.~(\ref{s-feqgb}), (\ref{nvagb}),
(\ref{theta}),  and  (\ref{Kgb}) yields the expansion parameter
\begin{equation}
\Theta = \frac{3}{2r} \left[1 + \frac{r^2}{4\alpha}\left[1 - \sqrt{1
+ 8\alpha \left(\frac{m(v)}{r^4}\right)}\right]\right]. \label{thgb}
\end{equation}
From the Eq.~(\ref{thgb}) it is clear that AH is the solution
of
\begin{equation}
\left[1 + \frac{r^2}{4\alpha}\left[1 - \sqrt{1 + 8\alpha
\left(\frac{m(v)}{r^4}\right)}\right]\right] = 0.
\end{equation}
i.e.,
\begin{equation}\label{aegb}
 r_{AH} = \sqrt{m(v) - 2\alpha}.
\end{equation}
In the relativistic limit $\alpha \rightarrow 0$ then $r_{AH}
\rightarrow \sqrt{m(v)}$. Hence our solution reduces to the solution \cite{mrm,gs} in 5D  space-time.
One sees that $g_{vv}( r_{AH} = \sqrt{m(v) - 2\alpha}) = 0$ implies that $r = \sqrt{m(v) - 2 \alpha}$
is time-like surface.  For an outgoing null geodesic $r = \sqrt{m(v) - 2 \alpha}$,
$\dot{r}$ is given by Eq.~(\ref{eq:de1gb}).  It is clear that presence of the  coupling constant
of  the Gauss-Bonnet terms $\alpha$ produces a change in the location of the AH.  Such a change could have a significant effect in the dynamical evolution of the black hole horizon.
\begin{equation}\label{rdd}
\ddot{r} = \frac{r\dot{r}}{4\alpha}\left(1 - \sqrt{1 + 8\alpha
\frac{m(v)}{r^4}}\right) + \frac{\frac{L}{2r^2} +  \frac{4
m(v)\dot{r}}{r^3}}{\sqrt{1 + 8\alpha\frac{m(v)}{r^4}}}
\end{equation}
At the time-like surface $r = \sqrt{m(v) - 2 \alpha}$, $\dot{r} = 0$
and $\ddot{r}> 0$ for $L >0$.  Hence the photon will escape from the $r=\sqrt{m(v)-2\alpha}$ and reach arbitrary large distance.

 On the other hand, The EH is a null three-surface which is the locus of outgoing
future-directed null geodesic rays that never manage to reach
arbitrarily large distances from the black hole and are determined
via Raychaudhuri equation. It can be seen to be equivalent to the
requirement that
\begin{equation}
\left[\frac{d^2r}{dv^2}\right]_{\mbox{EH}} \simeq ~ 0. \label{ehgb}
\end{equation}
An outgoing radial null geodesic satisfy
\begin{equation}
\frac{dr}{dv} = \frac{1}{2}\left[1 + \frac{r^2}{4\alpha}\left[1 -
\sqrt{1 + 8\alpha \left(\frac{m(v)}{r^4}\right)}\right]\right].
\label{rnggb}
\end{equation}
Then Eqs.~(\ref{Kgb}) and (\ref{thgb}) can be used to put
Eq.~(\ref{ehgb}) in the form
\begin{equation}
K \Theta_{EH} \simeq  \left[ \frac{3}{2r}\frac{\partial f}{\partial
v} \right]_{EH} \simeq  - \frac{3}{2{r_{EH}^3}} \frac{{L_M}}{\sqrt{1
+ 8\alpha \left(\frac{m(v)}{r_{EH}^4}\right)}},  \label{eh1gb}
\end{equation}
where the expansion is
\begin{equation}
\Theta_{EH}\simeq \frac{3}{2 r_{EH}}\left[1 +
\frac{r^2_{EH}}{4\alpha}\left[1 - \sqrt{1 + 8\alpha
\left(\frac{m(v)}{r^4_{EH}}\right)}\right]\right]. \label{thgb1}
\end{equation}
For the null vectors $l_a$ in Eq.~(\ref{nvagb}) and the component of
energy momentum tensor yields
\begin{equation}
R_{a b}l^{a}l^{b} =  \frac{3}{2r} \frac{\partial f}{\partial v}.
\label{chigb}
\end{equation}
The Raychaudhuri equation, with $\sigma = \omega = 0$ \cite{jygb}:
\begin{equation}
   \frac{d \Theta}{d v} = K \Theta - R_{ab}l^al^b- \frac{1}{2}
   \Theta^2.  \label{mregb}
\end{equation}
Thus neglecting $\Theta^2$, Eqs.~(\ref{eh1gb}), (\ref{chigb}) and
(\ref{mregb}), imply that
\begin{equation}
  \left[ \frac{d \Theta}{d v} \right]_{EH} \simeq 0.\label{eh}
\end{equation}
The event horizon in our case are therefore placed by
Eq.~(\ref{eh}).
Following \cite{rlm,mrm}, the solution can be immediately written
\begin{equation}
r_{EH} = \sqrt{m^{*}(v) - 2 \alpha},
\end{equation}
where
\begin{equation}
m^{*}(v) = m(v) - \frac{L}{K}.
\end{equation}
Thus the expression of the apparent horizon is exactly same as its counterpart Eq.~(\ref{aegb})
with the mass replaced by the effective mass $m^*.$   The region between the AH and the EH is defined as a
\emph{quantum ergosphere} \cite{jygb}.  All these results are consistent with known results
in 5D  space-time for the $\alpha \rightarrow 0$ case.
\section{Conclusion}
We have examined Vaidya radiating black-holes in Einstein-Gauss-Bonnet gravity.  The structure and location of the AH and EH are determined.
We have pointed out exact location of these horizons. It is clear that presence of the  coupling constant
of the Gauss-Bonnet terms $\alpha$ produces a change in the location of these horizons.  Such a change could have a significant effect in the dynamical evolution of the black hole horizon.  In particular, our results in the limit $\alpha \rightarrow 0$  reduced exactly  to  \emph{vis-$\grave{a}$-vis} 5D relativistic case.
\section*{Acknowledgments}
Authors would like to thank IUCAA, Pune for hospitality while this
work was done.

\end{document}